\documentclass[12pt]{iopart} 
\usepackage{graphicx}
\usepackage{citesort}
\usepackage{amssymb}
\usepackage{textcomp, libertine}
\usepackage{color}

\begin{document}

\title[Electron-phonon coupling and superconductivity in $\alpha$-MoB$_2$
       as a function of pressure]
       {Electron phonon coupling and superconductivity in $\alpha$-MoB$_2$
			 as a function of pressure}

\author{M-A Carmona-Galv\'an$^{1,3}$, R Heid$^2$, and O De la 
        Pe\~na-Seaman$^{3}$}

\address{$^1$ Facultad de Ciencias F\'isico Matem\'aticas (FCFM), Benem\'erita 
         Universidad Aut\'onoma de Puebla, Apartado Postal 1152 C.P. 72000 
				 Puebla, Puebla, M\'exico}
\address{$^2$ Institut f\"{u}r QuantenMaterialien und Technologien, 
         Karlsruher Institut f\"{u}r Technologie (KIT), D-76021 Karlsruhe, 
				 Germany}
\address{$^3$ Instituto de F\'isica ``Ing. Luis Rivera Terrazas'', Benem\'erita 
         Universidad Aut\'onoma de Puebla, Av. San Claudio \& Blvd. 18 Sur, 
				 Ciudad Universitaria, C.P. 72570, Puebla, Puebla, M\'exico}

\eads{oseaman@ifuap.buap.mx}

\begin{abstract}
We have studied the lattice dynamics, electron-phonon coupling, and 
superconducting properties of $\alpha$-MoB$_2$, as a function of applied 
pressure, within the framework of density functional perturbation theory using 
a mixed-basis pseudopotential method.
We found that phonon modes located along the A$-$H, H$-$L, and L$-$A 
high-symmetry paths exhibit large phonon linewidths and contribute 
significantly to the electron-phonon coupling constant. 
Although linewidths are particularly large for the highest-frequency optical 
phonon modes (dominated by B vibrations), their contribution to the 
electron-phonon coupling constant is marginal. The latter is largely controlled 
by the acoustic low-frequency modes of predominantly Mo character. 
It was observed that at a pressure of $90$~GPa, where $\alpha$-MoB$_2$ forms, the 
phonon-mediated pairing falls into the strong-coupling regime, and the estimate 
for the superconducting critical temperature $T_c$ agrees well with experimental 
observations. When further increasing the applied pressure, a reduction of $T_c$ is 
predicted, which correlates with a hardening of the acoustic low-frequency 
phonon modes and a decrease of the electron-phonon coupling parameter.
\end{abstract}

\vspace{2pc}

\submitto{PHYSSCR}

\section{Introduction}

The discovery of superconductivity in MgB$_2$ more than twenty 20 years ago 
\cite{naga}, with a critical temperature of $T_c \approx 39$~K, energized the 
search for new superconducting materials within the family of diborides. Such 
a quest was pursued almost immediately after its discovery both experimentally 
and computationally \cite{buze}. 
After several years of research, the conclusion was reached that MgB$_2$ is 
already optimized by nature, in the sense that attempts to improve its 
superconducting properties by doping \cite{kaza,karp,dagh,omar1} or pressure 
\cite{monte,wang} always resulted in a reduction of $T_c$ in comparison with 
MgB$_2$, or even in a non-superconducting material, like the sibling system 
AlB$_2$ \cite{renk}. 

Transition-metal diborides constitute an important sub-class in this context. 
A typical example studied was NbB$_2$ with a wide range of measured $T_c$ from 
$0.62$~K to $9$~K \cite{leya,kote,take}. MoB$_2$ attracted attention as well. 
While it is not a superconductor in its pristine form, superconductivity can be 
induced by substitution of $4$\% Zr with a $T_c \approx 6$~K \cite{muzzy}. It 
was not until 2022 that the discovery of superconductivity in MoB$_2$ under 
applied pressure was reported \cite{pei}. At an applied pressure of 
approximately $20$~GPa, MoB$_2$ becomes superconducting with a very low $T_c$ 
of less than $2$~K. At these pressures, MoB$_2$ takes a rhombohedral crystal 
structure (space group $R\bar{3}m$), known also as $\beta$-MoB$_2$. $T_c$ 
rapidly increases as a function of pressure, reaching $T_c \approx 27$~K at a 
pressure of $p_c\approx 70$~GPa, where it gradually transforms into the hexagonal 
$\alpha$-MoB$_2$ structure (space group $P6/mmm$, $D_{6h}^1$ no.$191$ 
\cite{jon}). With further increase of pressure, $\alpha$-MoB$_2$ experiences a 
less dramatical $T_c$ increase, which culminates at $110$~GPa in a maximum $T_c$ 
of $32.4$~K \cite{pei}. 

Theoretical calculations have suggested that the mechanism for such a high 
$T_c$ value in $\alpha$-MoB$_2$ is quite different from the one in MgB$_2$. In 
particular, while for MgB$_2$ the pairing is coming from the strong coupling 
between the $\sigma$-bands and the B-related $E_{2g}$ phonon modes 
\cite{kortus,floris,bohn,geerk}, in MoB$_2$ the pairing involves electronic 
states of the Mo-$d$ character and a combination of Mo-related low-frequency 
phonon modes with B-dominated ones \cite{pei,quan}. In fact, Quan \etal 
\cite{quan} concluded that the source of the MoB$_2$ $T_c$ is the so called 
electron-displaced atom scattering factor $I^2$, which is closely related to 
the electron-phonon (e-ph) matrix elements of the Eliashberg theory 
\cite{eliash} (see equation \ref{eq03}). However, a detailed analysis about how 
this factor and other ingredients involved in conventional superconductivity 
(like phonon frequencies, linewidths, or electron-phonon coupling parameter) 
are evolving as a function of pressure is lacking. 

In this paper we present a thorough study of the lattice dynamics, 
electron-phonon coupling, and superconducting $T_c$ of  $\alpha$-MoB$_2$ as 
a function of applied pressure, from $70$~GPa to $300$~GPa, within the 
framework of density functional theory (DFT) \cite{kohn} and density functional 
perturbation theory (DFPT) \cite{louie,baro,bohn1,bohn2} using a mixed-basis 
pseudopotential method \cite{mbpp}. Superconducting properties are analyzed 
in the framework of the Eliashberg theory \cite{eliash}. 
We give a detailed description of the phonon linewidths and electron-phonon 
coupling as a function of applied pressure. In particular, we analyze the 
contributions of different phonon modes to these quantities, and determine its 
specific role for inducing the high $T_c$ value of $\alpha$-MoB$_2$. 
For comparison, we also present a similar analysis for the sibling 
system NbB$_2$, which is a low-$T_c$ superconductor with intermediate coupling.
The paper is organized as follows. In section 2 we describe the computational 
details of our calculations. The results for the evolution of lattice dynamics, 
e-ph coupling and $T_c$ as a function of pressure are presented in section 3. 
Finally, in section 4 the main findings are summarized.

\section{Computational details}

The present density-functional calculations \cite{kohn} were performed with the 
mixed-basis pseudopotential method (MBPP) \cite{mbpp}. Norm-conserving 
pseudopotentials for Mo, Nb, and B were generated according to the Vanderbilt 
description \cite{van} and include partial-core correction. For Mo and Nb, 
semicore $4s$ and $4p$ states were taken into the valence space.
The current method applies a mixed-basis scheme, which uses a combination of 
local functions and plane waves for the representation of the valence states. 
We used $s$, $p$, and $d$-type functions for Mo and Nb, while for B only $s$ 
and $p$-type, supplemented by plane waves up to a kinetic energy of 
$32$~Ry. Present calculations were performed with the PBE \cite{perdew} form of 
the GGA exchange-correlation functional. The Monkhorst-Pack special $k$-point 
sets technique, with a Gaussian smearing of $0.25$~eV and a grid of 
$18\times18\times18$, was used for the the Brillouin-zone integration.
Phonon properties are calculated via density functional perturbation theory 
(DFPT) \cite{louie,baro} as implemented in the MBPP code \cite{bohn1,bohn2}. 
The phonon dispersions are obtained by a Fourier interpolation of dynamical 
matrices calculated on a $6\times6\times6$ $q$-point mesh. For the calculation 
of e-ph coupling matrix elements, a denser $36\times 36\times 36$ $k$-point 
mesh was necessary. 

Through the knowledge of the phonon dispersion and e-ph matrix elements the 
Eliashberg function is accessible, 

\begin{equation}
\alpha^2F(\omega)=\frac{1}{2\pi\hbar N(E_F)}\sum_{{\bf q}\eta}
\frac{\gamma_{{\bf q}\eta}}{\omega_{{\bf q}\eta}}
\delta(\omega-\omega_{{\bf q}\eta}),
\label{eq01}
\end{equation}

\noindent
with $N(E_F)$ as the electronic density of states at the Fermi level, per atom 
and spin; $\omega_{{\bf q}\eta}$ as the frequency of the phonon mode at the 
$\bf{q}$-vector and branch $\eta$, and the phonon linewidths 
$\gamma_{{\bf q}\eta}$ given by

\begin{equation}
\gamma_{{\bf q}\eta}=2\pi\omega_{{\bf q}\eta}\sum_{{\bf k}\nu \nu'}
\left|  g^{{\bf q}\eta}_{{\bf k}+{\bf q}\nu',{\bf k}\nu} \right|^2
\delta(\epsilon_{{\bf k}\nu}-E_F)\delta(\epsilon_{{\bf k}+{\bf q}\nu'}-E_F),
\label{eq02}
\end{equation}

\noindent
where $\epsilon_{{\bf k}\nu}$ is the one-electron band energy with momentum 
${\bf k}$ and band index $\nu$. In the last equation, 
$g^{{\bf q}\eta}_{{\bf k}+{\bf q}\nu',{\bf k}\nu}$ represents the coupling 
matrix element for scattering of an electron from a ${\bf k}\nu$ electronic 
state to another ${\bf k}+{\bf q}\nu'$ state, by a phonon ${\bf q}\eta$, and 
is given by

\begin{equation}
g^{{\bf q}\eta}_{{\bf k}+{\bf q}\nu',{\bf k}\nu}= \sqrt{\frac{\hbar}
{2\omega_{{\bf q}\eta}}}\sum_{{\kappa a}}\frac{1}{\sqrt{M_{\kappa}}}
\eta^{{\bf q}\eta}_{\kappa a}\left\langle {\bf k}+{\bf q}\nu' \left|  
\delta^{{\bf q}}_{\kappa a} V \right| {\bf k}\nu \right\rangle,
\label{eq03}
\end{equation} 

\noindent
with $M_\kappa$ as the mass of the $\kappa$-th atom in the unit cell, and 
$\eta^{{\bf q}\eta}_{\kappa a}$ as the normalized eigenvector of the 
corresponding phonon mode ${\bf q}\eta$. The quantity 
$\delta^{{\bf q}}_{\kappa a} V$ represents the first-order change of the 
total crystal potential, with respect to the displacement of the $\kappa$-th 
atom in the $a$ direction. 

From $\alpha^2F(\omega)$ we can obtain some useful integrated quantities, like 
the average Allen-Dynes characteristic phonon frequency $\omega_{log}$

\begin{equation}
\omega_{\rm log}=\exp \left( \frac{2}{\lambda} \int^{\infty}_{0}
d{\omega}\frac{\ln(\omega)}{\omega} \alpha^{2}F(\omega) \right)\,,
\label{eq04}
\end{equation}

\noindent
the square-average phonon frequency $\bar{\omega}_2$

\begin{equation}
\bar{\omega}_2=\left\langle\omega^2\right\rangle^{1/2}=
\left(\frac{2}{\lambda} \int^{\infty}_{0}
d{\omega}\alpha^{2}F(\omega)\omega\right)^{1/2},
\label{eq05}
\end{equation}

\noindent
the average e-ph coupling constant $\lambda$

\begin{equation}
\lambda=2\int^{\infty}_{0} \frac{d\omega}{\omega} \alpha^2F(\omega)=
\frac{1}{\pi \hbar N(E_F)}\sum_{{\bf q}\eta}
\frac{\gamma_{{\bf q}\eta}}{\omega^2_{{\bf q}\eta}}  \,,
\label{eq06}
\end{equation}

\noindent
as well as the frequency-dependent $\lambda$, given by:

\begin{equation}
\lambda(\omega)=2\int^{\omega}_{0} \frac{d\omega'}{\omega'} 
\alpha^2F(\omega')\,.
\label{eq07}
\end{equation}

Finally, $\alpha^2F(\omega)$ is used to determinate the superconducting 
critical temperature, $T_c$, by solving the Eliashberg gap equations 
\cite{eliash,marsi} numerically.

\section{Results and discussion}

\begin{figure}[htbp]
\centering
\vspace*{0.0 true cm}
\includegraphics*[scale=0.46,trim=0mm 4cm 0mm 0.5cm,clip]{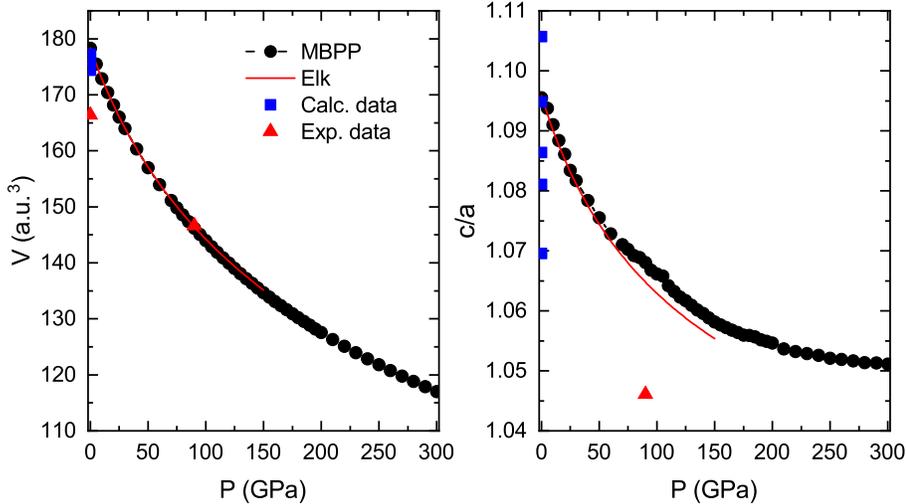}
\caption{\label{fig01} Calculated $p(V)$ equation of state and optimized $c/a$ 
parameter, as a function of applied pressure for $\alpha$-MoB$_2$ obtained by 
two different band-structure methods (MBPP \cite{mbpp} and Elk \cite{elk}), 
compared with experimental data \cite{pei,tao} (red triangles), and calculated 
results reported previously \cite{qi,ding,xu,shein,deli} (blue squares).}
\end{figure}

The $\alpha$-MoB$_2$ has an AlB$_2$-type structure, consisting of planar
close packed layers of Mo (at $(0,0,0)$, $1a$ Wyckoff position) and B atoms (at 
$(2/3,1/3,1/2)$, $2d$ Wyckoff position) alternated along the $c$-axis of the 
hexagonal unit cell \cite{pei,tao,jon}. Its structure was fully optimized by 
energy minimization, that is, for each fixed $V$ the $c/a$ parameter was 
optimized in order to get the $E(V)$ and $p(V)$ equations of state (see 
\fref{fig01}).


The current results are compared with available experimental data \cite{pei,tao}, 
as well as reported calculated values \cite{qi,ding,xu,shein,deli}. Our results 
are in remarkable agreement with the data of Pei \etal \cite{pei,tao} at 
$90$~GPa for both, the volume (a difference of around $0.3$\%) and also the $c/a$ 
ratio (difference of $2.1$\%). In addition, structure-optimization calculations 
were also performed with the full-potential Elk code \cite{elk}, showing an 
excellent agreement with the MBPP-code calculations, which demonstrates the high 
accuracy of the constructed pseudopotentials. 

\begin{figure*}[htbp]
\centering
\vspace*{0.0 true cm}
\includegraphics*[scale=0.54,trim=0mm 12.5cm 0mm 0.5cm,clip]{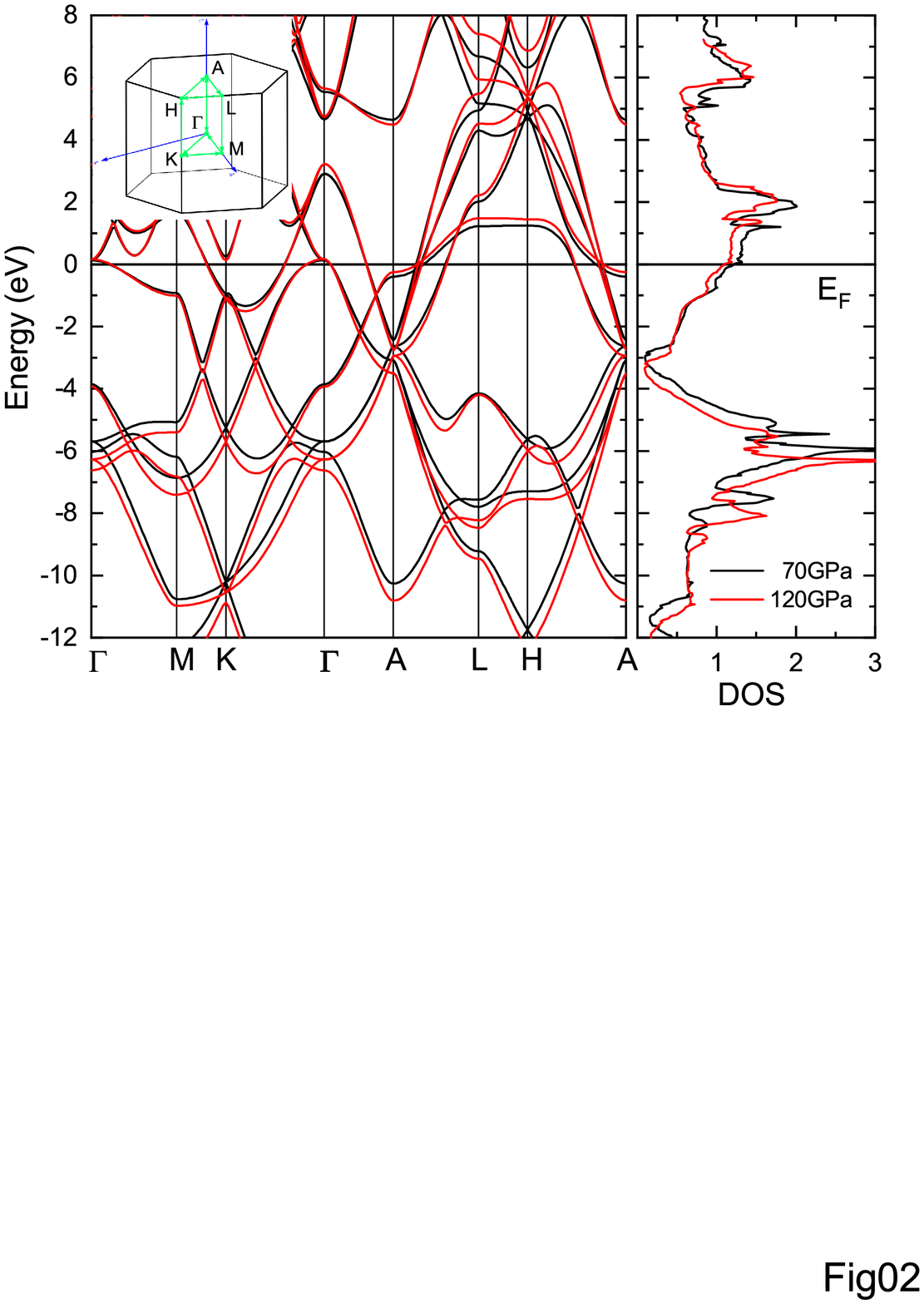}
\caption{\label{fig02} Comparison of the electronic band structure and 
density of states (DOS) for $70$~GPa and $120$~GPa. The first Brillouin zone 
for the hexagonal structure is presented as inset (generated by XCrySDen) 
\cite{hex}.}
\end{figure*}

The comparison of the electronic band structure and density of states for 
two different pressure values, $70$~GPa and $120$~GPa, is presented in 
\fref{fig02}. As already pointed out previously \cite{quan}, the bands around 
the Fermi level ($E_F$) are dominated by the Mo $d$ states, with a very minor 
participation of B states. This property is maintained across the whole pressure 
range, and results in very modest changes around the Fermi level. The main 
pressure effects are a band-width increase and a small reduction of the states 
at the Fermi level ($N(E_F)$).


\begin{figure*}[htbp]
\centering
\vspace*{0.0 true cm}
\includegraphics*[scale=0.58,trim=0mm 2.0cm 0mm 2.0cm,clip]{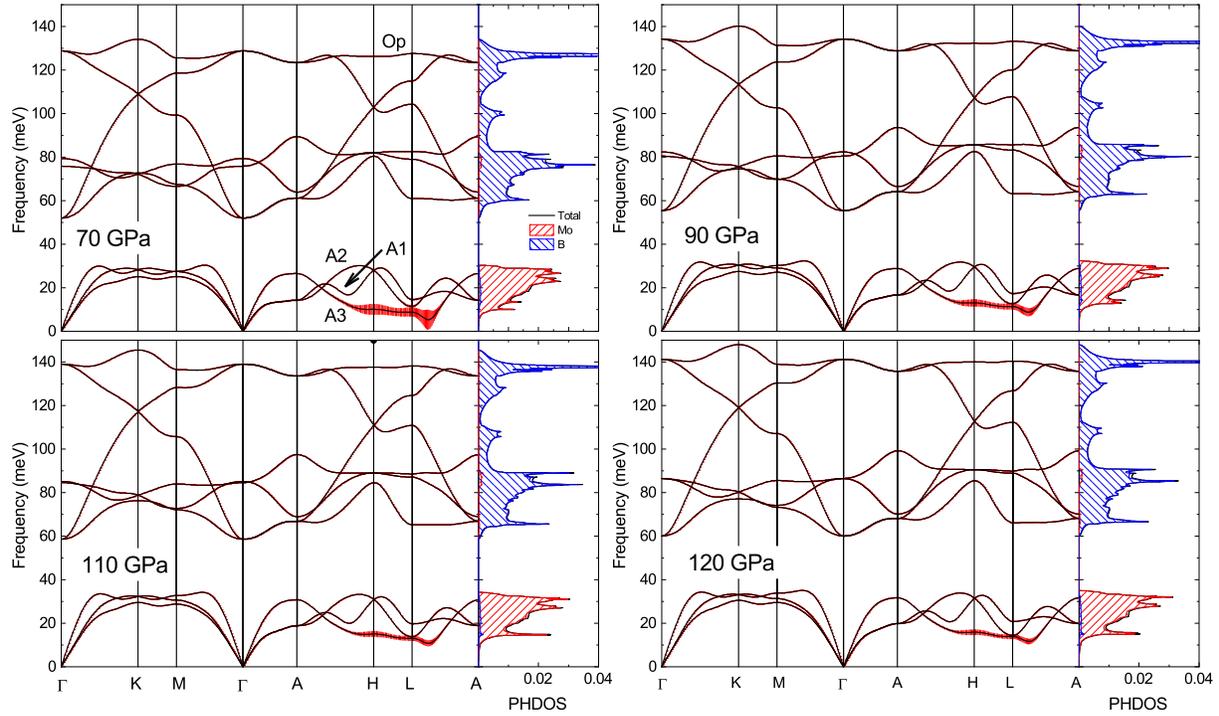}
\caption{\label{fig03} Phonon dispersions and phonon density of states (PHDOS)
for $\alpha$-MoB$_2$, calculated at selected pressures: $70$~GPa, $90$~GPa, 
$110$~GPa, and $120$~GPa. Vertical red lines correspond to the e-ph coupling 
constant $\lambda_{\bf{q}\eta}$.
The labels correspond to the acoustic phonon branches (A$1$, A$2$, and A$3$), 
as well as the highest-optic one (Op) at the A$-$H$-$L$-$A paths.}
\end{figure*}

In \fref{fig03} the phonon dispersion along high-symmetry directions 
as well as its corresponding phonon density of states (PHDOS), for 
specific applied pressure values, are presented. The chosen pressures span 
across the stability region of the $\alpha$-MoB$_2$ structure \cite{pei}. The 
main characteristics of the phonon spectrum, as previously observed 
\cite{pei,quan}, are found for the whole pressure range. On the one hand, the 
low-frequency region dominated by Mo vibrations; the high-frequency one ruled 
by B modes; and the frequency gap that separates them. On the other hand, the 
acoustic low-frequency modes along the L-A path, which exhibit a phonon anomaly 
close to L-point, as well as the soft acoustic branches along the A-H and H-L 
paths. Interestingly, the acoustic mode with lowest frequency (labeled as A$3$) 
is the one with the largest e-ph coupling constant contribution, given 
by the red vertical lines in \fref{fig03}.
In general, the main effect of the applied pressure on the phonon spectra is a 
generalized hardening of the phonon frequencies, which directly weakens the 
observed phonon anomaly at the L-A path, and reduces at the same time its 
strong e-ph contribution.


\begin{figure*}[htbp]
\centering
\vspace*{0.0 true cm}
\includegraphics*[scale=0.82,trim=0mm 20.0cm 0mm 0mm,clip]{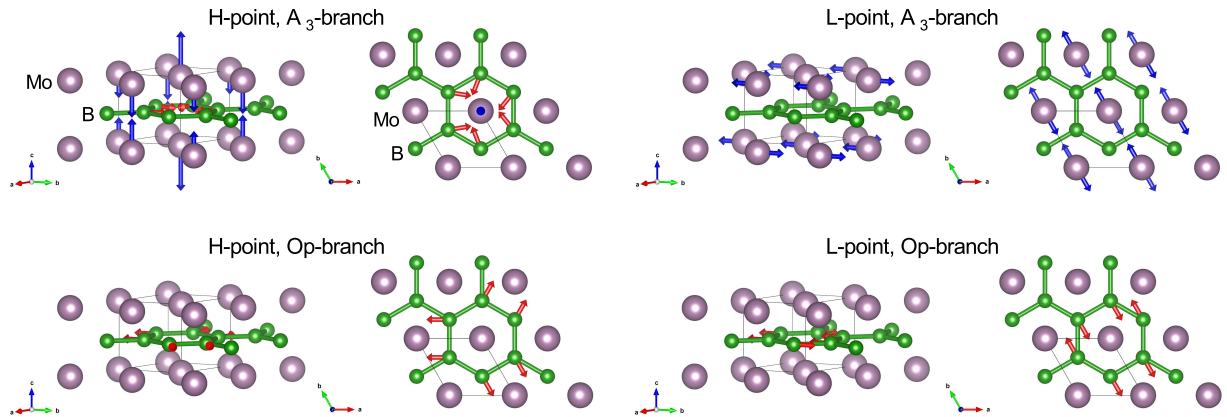}
\caption{\label{fig04} Phonon modes corresponding to the acoustical A$3$ 
and optical Op branches at the H- and L-points \cite{vesta}. The size of the 
arrows indicate the vibrational directions and the corresponding magnitude. For 
the case of the H-point A$3$-branch, the B related arrows where scaled ($4$X) in 
order to be noticeable.}
\end{figure*}


In addition, we also present the phonon modes corresponding to the 
acoustical A$3$ and optical Op branches at the H- and L-points of the IBZ
(\fref{fig04}). The displacement patterns show that the A$3$ related modes are 
clearly dominated by Mo atoms, while the Op ones are ruled completely by the B 
atoms.


\begin{figure*}[htbp]
\centering
\vspace*{0.0 true cm}
\includegraphics*[scale=0.50,trim=0mm 9.0cm 0mm 1.0cm,clip]{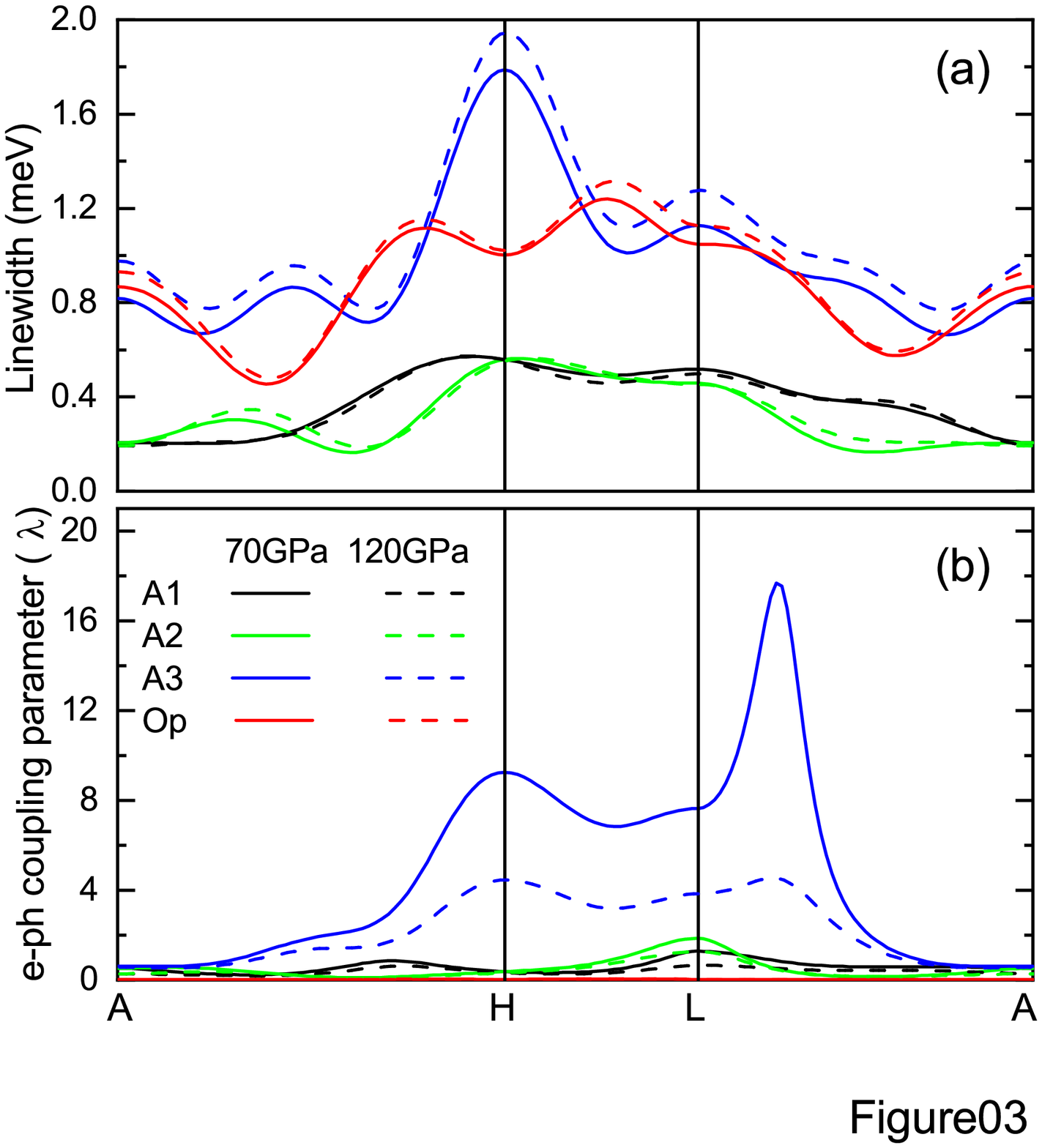}
\caption{\label{fig05} (a) Linewidths and (b) e-ph coupling constant for 
$\alpha$-MoB$_2$, at $70$~GPa (solid lines) and $120$~GPa (dashed lines) along 
the A$-$H$-$L$-$A paths, for the three acoustic branches (A$1$, A$2$, and A$3$) 
and the highest-frequency optical branch (Op).}
\end{figure*}


A closer inspection of the individual phonon linewidths and mode couplings 
revealed that important contributions are attributed to the acoustic 
phonon branches (A$1$, A$2$, and A$3$) and the highest optic one (Op), in 
particular along at the A$-$H$-$L$-$A paths (\fref{fig03}).
In \fref{fig05}, linewidths and e-ph coupling constants of these modes are show 
along these high-symmetry directions for two pressures. The largest linewidths 
(\fref{fig05}a) are found for the A$3$ branch, with a particular strong peak 
located at the H-point, followed closely by the Op branch. These results 
indicate an important participation of phonon modes dominated by Mo (A$3$) and 
also by B (Op) in the e-ph coupling (equation \ref{eq02}) reflected by the 
phonon linewidths (equation \ref{eq03}). 
However, for the e-ph coupling constants shown in \fref{fig05}b, the influence 
of B phonon-modes is faded away due to the factor $1/\omega_{\bf{q}\eta}^2$ 
entering its definition (equation \ref{eq06}). In contrast, the large e-ph 
coupling constants of the acoustic branch A$3$ is boosted by the low 
frequencies of these Mo phonon modes, especially around the phonon anomaly 
close to the L-point. With increasing pressure, while the linewidths increase a 
little bit, $\lambda_{\bf{q}\eta}$ strongly reduces, correlating with the 
observed hardening of this acoustic branch.

In order to analyze the evolution of the superconducting properties as a 
function of pressure, the Eliashberg function $\alpha^2F(\omega)$, the 
e-ph coupling constant $\lambda$, the Allen-Dynes characteristic phonon 
frequency $\omega_{log}$, and the square-average phonon frequency 
$\bar{\omega}_2$ were calculated for each case.


\begin{figure*}[htbp]
\centering
\vspace*{0.0 true cm}
\includegraphics*[scale=0.62,trim=0mm 1.5cm 0mm 0mm,clip]{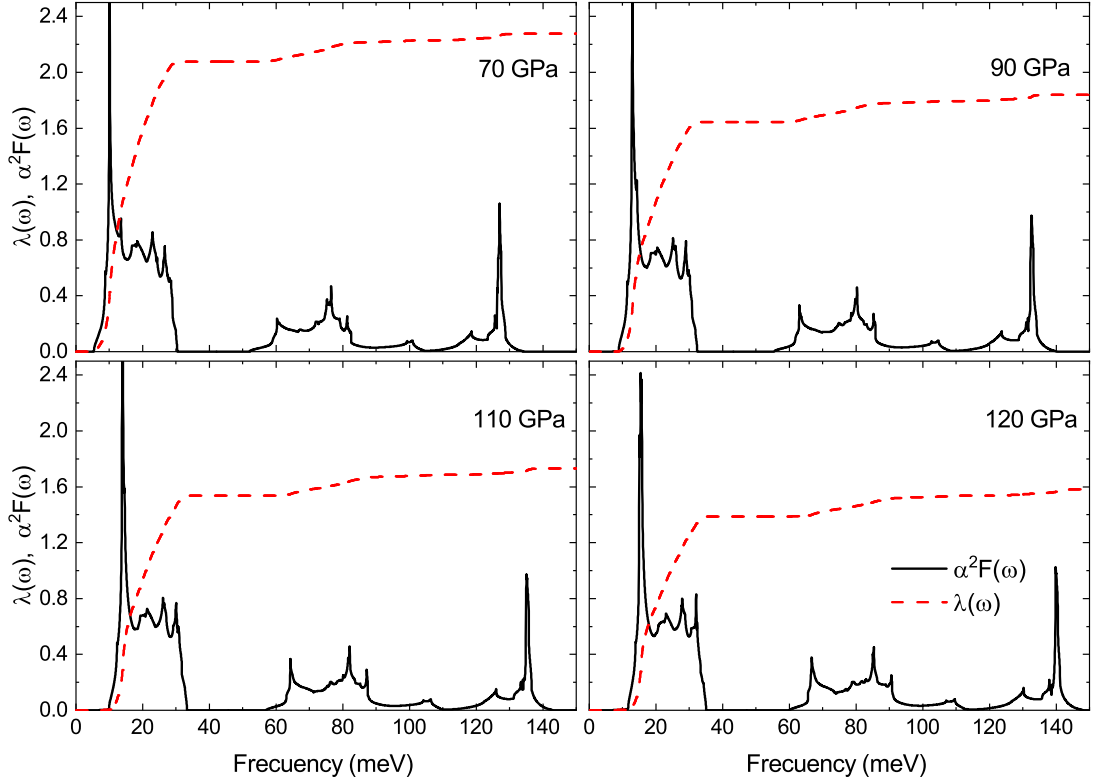}
\caption{\label{fig06} $\alpha$-MoB$_2$ Eliashberg function $\alpha^2F(\omega)$ 
(black solid line) and frequency-dependent e-ph coupling constant 
$\lambda(\omega)$ (red dashed line) for specific applied pressure values.}
\end{figure*}


The Eliashberg functions for selected pressures are presented in \fref{fig06}, 
together with $\lambda(\omega)$. In all cases, the largest contribution for 
$\alpha^2F(\omega)$ and $\lambda(\omega)$ comes from the acoustic low-frequency 
region, dominated almost completely by Mo phonon modes along the A-H, H-L, and 
specially the L-A paths, where the phonon anomaly is located. As expected, the 
largest coupling corresponds to the pressure that is closest to the phase 
transition: $70$~GPa with $\lambda \approx 2.3$. As pressure increases, the 
coupling reduces, at the same time that the observed phonon anomaly attenuates, 
which is a direct consequence of the general hardening of the phonon spectrum,
as previously discussed.


\begin{figure*}[htbp]
\centering
\vspace*{0.0 true cm}
\includegraphics*[scale=0.60,trim=0mm 2.0cm 30mm 0mm,clip]{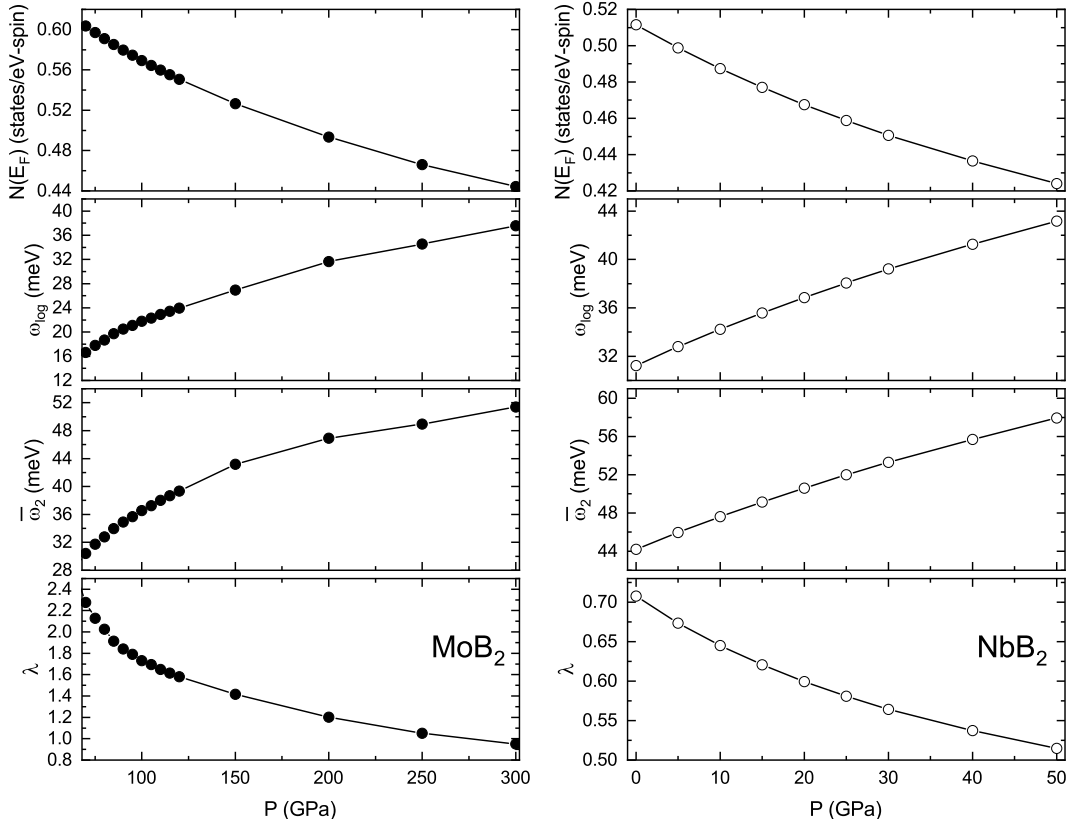}
\caption{\label{fig07} Density of states at the Fermi level ($N(E_F)$), the 
Allen-Dynes characteristic phonon frequency ($\omega_{log}$), the 
square-average phonon frequency ($\bar{\omega}_2$), and the e-ph coupling 
constant ($\lambda$), as a function of applied pressure, for $\alpha$-MoB$_2$ 
(left) and NbB$_2$ (right).} 
\end{figure*}


The evolution of the coupling related quantities, namely the density of states 
at the Fermi level ($N(E_F)$), $\omega_{log}$, $\bar{\omega}_2$, and $\lambda$, 
as a function of pressure, are presented in \fref{fig07}. There is a nice 
agreement of these quantities with the reported values in literature at 
$90$~GPa \cite{pei,quan}, although our calculated $\lambda=1.84$ is slightly 
larger (between $10$\% and $15$\%). This can be due to the slight difference on 
the structural parameters (see \fref{fig01}) or pseudopotential construction. 
From the evolution of $\lambda$, it can be seen that the strong pressure 
dependence of the coupling is coming mainly from the low-frequency phonons 
(traced by $\omega_{log}$ and $\bar{\omega}_2$), while $N(E_f)$ does not 
exhibit dramatic changes as a function of pressure. $\alpha$-MoB$_2$ remains in 
the strong-coupling regime until $300$~GPa, where $\lambda = 0.95$, while 
$\omega_{log} = 37.58$~meV, and $\bar{\omega}_2 = 51.39$~meV. 

For comparison, we also calculated the same e-ph parameters, as a function of 
applied pressure, for the sibling compound NbB$_2$ (with the same crystal 
structure) at its own optimized structural parameters (see \fref{fig07}). 
NbB$_2$ was studied more or less at the same time when superconductivity in 
MgB$_2$ was discovered. This was done with the idea to find related materials 
with improved superconducting properties. It turned out, however, that NbB$_2$ 
has an intermediate coupling ($\lambda=0.67$) and a low $T_c$ value (approx. 
$8.4$~K) \cite{heid}. 
Although NbB$_2$ has lower $\lambda$ values than MoB$_2$ (the highest 
calculated $\lambda$ for NbB$_2$ is $0.71$ at $p=0$~GPa), the trends as a 
function of pressure for the coupling-related quantities are basically the 
same: a reduction of $N(E_F)$, a phonon hardening, and a $\lambda$ decrease.


\begin{figure*}[htbp]
\centering
\vspace*{0.0 true cm}
\includegraphics*[scale=0.60,trim=0cm 8.5cm 0cm 0cm,clip]{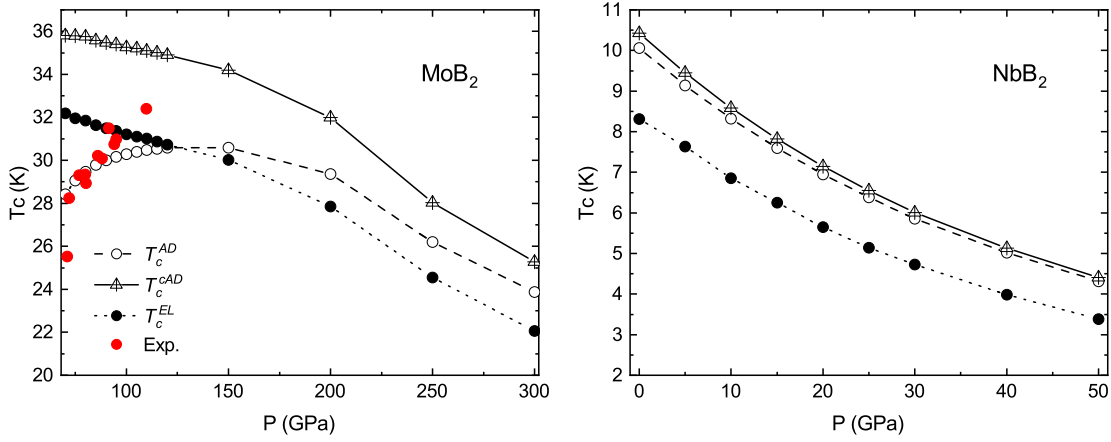}
\caption{\label{fig08} Evolution of $T_c$ as a function of pressure for 
$\alpha$-MoB$_2$ (left) and NbB$_2$ (right) using the standard Allen-Dynes 
\cite{allen} equation ($T^{AD}_c$), the corrected Allen-Dynes equation 
($T^{cAD}_c$) \cite{allen}, and the numerical solution of the isotropic 
Eliashberg gap equations ($T^{EL}_c$) \cite{eliash,marsi}. Comparison with 
experimental results available in literature for $\alpha$-MoB$_2$ \cite{pei} 
(red dots).}
\end{figure*}


In order to analyze the evolution of $T_c$ as a function of pressure, we 
applied three different schemes: $(1)$, the standard Allen-Dynes 
equation \cite{allen},

\begin{equation}
T^{AD}_c = \frac{\omega_{log}}{1.20}\mbox{exp}\left(-\frac{1.04(1+\lambda)}
{\lambda-\mu^*(1+0.62\lambda)}\right),
\label{eq07}
\end{equation} 

\noindent
$(2)$, the corrected Allen-Dynes equation for strong-coupling systems (normally 
for $\lambda \leq 1.3$) \cite{allen},

\begin{equation}
T^{cAD}_c = \frac{f_1f_2\omega_{log}}{1.20}\mbox{exp}\left(-\frac{1.04(1+\lambda)}
{\lambda-\mu^*(1+0.62\lambda)}\right),
\label{eq08}
\end{equation} 

\noindent
where the correction factors to describe the strong-coupling regime are

\begin{eqnarray}
f_1 = \left[1+(\lambda/\Delta_1)^{3/2}\right]^{1/3}, \label{eq09}\\
f_2 = 1 + \frac{(\bar{\omega}_2/\omega_{log}-1)\lambda^2}
{\lambda^2+\Delta_2^2}, \label{eq10}
\end{eqnarray}

\noindent
and the parameters $\Delta_1$ and $\Delta_2$ given by

\begin{eqnarray}
\Delta_1 = 2.46(1+3.8\mu^*), \label{eq11}\\
\Delta_2 = 1.82(1+6.3\mu^*)\left(\bar{\omega}_2/\omega_{log}\right), 
\label{eq12}
\end{eqnarray}

\noindent
and finally $(3)$, by solving the isotropic Eliashberg gap equations 
\cite{eliash,marsi} numerically, $T^{EL}_c$, using the calculated 
$\alpha^2F(\omega)$ for each considered pressure.

\noindent
Results obtained for the three schemes, for both $\alpha$-MoB$_2$ and NbB$_2$, 
are presented in \fref{fig08}, using in all cases the same Coulomb 
pseudopotential parameter $\mu^*=0.13$, in order to be as close as possible to 
the previously reported $T_c$ values for $p=90$~GPa \cite{pei,quan}. 
As expected, there are quantitative differences between the $T_c$ estimates, in 
particular for the low-pressure region, where $\alpha$-MoB$_2$ is in the 
strong-coupling regime.
While both strong-coupling schemes ($T^{cAD}_c$ and $T^{EL}_c$) predict a 
monotonous superconducting temperature reduction as a function of pressure, 
$T^{AD}_c$ first increases slightly from $70$~GPa to approximately $150$~GPa, 
followed by a decrease. For $p > 250$~GPa, $T^{AD}_c$ and $T^{cAD}_c$ are 
getting closer, a clear indication of the transition to a more moderate 
coupling region. 
For NbB$_2$, all three $T_c$ estimates reveal the same pressure dependence, 
while $T^{AD}_c$ and $T^{cAD}_c$ agree almost quantitatively. This behavior is 
expected, since NbB$_2$ has an e-ph coupling that goes from intermediate to low 
coupling strength, as applied pressure increases.
From these results it is clear that the use of $T^{AD}_c$ is not adequate for 
a strong-coupling system like $\alpha$-MoB$_2$, showing misleading values and 
even wrong tendencies, as noted previously \cite{quan}. 
A possible reason of the apparent disagreement between our calculated 
$T_c$ with experimental data \cite{pei} is that, very likely, the measured 
MoB$_2$ samples below $p=90$~GPa possess a different crystal structure, 
or consist of a mix of different phases, as mentioned by the authors of the 
experimental work.
However, for pressures at (or above) $90$~GPa, our calculated $T^{EL}_c$ (by 
solving the Eliashberg gap equations) are around $\pm$ $1$~K from the reported 
measurements and, interestingly, $T^{EL}_c$ shows the best agreement with the 
reported experimental data at $90$~GPa. 
We note that, within the framework of the Eliashberg theory, solving the (isotropic) 
gap equations with $\alpha^2F(\omega)$ as input is the most direct way to calculate 
the superconducting temperature, and is superior to the other two approaches, which 
only provide approximations to its solution.
Such a $T_c$ reduction as a function of applied pressure, as obtained from our 
calculations for $\alpha$-MoB$_2$ and NbB$_2$, is also observed experimentally 
for Nb-substituted MoB$_2$ (Nb$_{0.25}$Mo$_{0.75}$B$_2$) \cite{lim}. There, a 
steady $T_c$ reduction is reported from $8$~K at $0$~GPa to $4$~K at $50$~GPa, 
followed by a gradual rise to $5.5$~K at $170$~GPa that is accompanied by a 
significant broadening of the superconducting transition width \cite{lim}.

\section{Conclusions}

To summarize, we have performed a first-principles linear-response study of the 
lattice dynamical properties, electron-phonon coupling, and superconductivity 
of $\alpha$-MoB$_2$ as a function of applied pressure (from $70$~GPa to 
$300$~GPa). We found that the electron-phonon interaction induces large phonon 
linewidths for modes located specifically along the A$-$H, H$-$L, and L$-$A 
high-symmetry paths, where a phonon anomaly is present. The largest linewidths 
are displayed by the highest-frequency optical phonon mode (ruled by B 
vibrations) and the acoustic low-frequency phonon modes (involving mainly Mo 
atoms). However, the contribution of the optical phonon mode to the 
electron-phonon coupling constant is diminished because of its high-frequency 
value, while the dominating one is coming from the lowest-frequency acoustic 
phonon mode. 
As pressure increases, the phonon spectrum hardens, in particular the acoustic 
low-frequency phonon modes, and the electron-phonon coupling constant 
decreases, while the density of states at the Fermi level barely changes. 
Estimates for $T_c$, obtained either way by the corrected Allen-Dynes equation 
or by solving the Eliashberg gap equations, show a decrease as a function of 
applied pressure, which correlates with the phonon hardening and the reduction 
of $\lambda$. 
We found a good agreement between the experimental $T_c$ values and the 
calculated ones for $90$~GPa and $110$~GPa. However, data for larger applied 
pressure values are needed to allow a more complete assessment of the predicted 
tendencies of $T_c$ for $\alpha$-MoB$_2$.

\ack{}
This research was partially supported by the Consejo Nacional de Humanidades, 
Ciencias y Tecnolog\'ias (CONAHCYT, M\'exico) under Grant No. FOP16-2021-01-320399; 
Vicerrector{\'\i}a de Investigaci\'on (VIEP), Benem\'erita Universidad 
Aut\'onoma de Puebla (BUAP) under Grant No. 100517450-VIEP2023; and the 
Karlsruher Institut f\"{u}r Technologie (KIT), Germany.

\section*{References}
\bibliography{mob2_press_arkiv_v02}

\end{document}